%% file: paper.tex
\documentclass[submission, Phys]{SciPost}

\input{incl_settings.tex} 
\input{incl_shortcuts.tex}
\graphicspath{{./figs/}}

\begin{document}


\begin{center}{\Large \textbf{
The \madnis Reloaded 
}}\end{center}

\begin{center}
Theo Heimel\textsuperscript{1},
Nathan Huetsch\textsuperscript{1},
Fabio Maltoni\textsuperscript{2,3},\\
Olivier Mattelaer\textsuperscript{2}, 
Tilman Plehn\textsuperscript{1},
and Ramon Winterhalder\textsuperscript{2}
\end{center}

\begin{center}
{\bf 1} Institut f\"ur Theoretische Physik, Universit\"at Heidelberg, Germany
\\
{\bf 2} CP3, Universit\'e catholique de Louvain, Louvain-la-Neuve, Belgium
\\
{\bf 3} Dipartimento di Fisica e Astronomia, Universit\'a di Bologna, Italy
\end{center}

\begin{center}
\today
\end{center}


\section*{Abstract}
{\bf In pursuit of precise and fast theory predictions for the LHC, we
  present an implementation of the \madnis method in the \textsc{MadGraph}
  event generator. A series of improvements in \madnis further enhance 
  its efficiency and speed. We validate this
  implementation for realistic partonic processes and find
  significant gains from using modern machine learning in
  event generators.}

\vspace{10pt}
\noindent\rule{\textwidth}{1pt}
\tableofcontents\thispagestyle{fancy}
\noindent\rule{\textwidth}{1pt}
\vspace{10pt}

\clearpage
\section{Introduction}
\label{sec:intro}

One of the defining aspects of the precision-LHC program is that we
compare vast amounts of scattering data with first-principle
predictions. These predictions come from multi-purpose LHC event
generators, for instance \pythia~\cite{Sjostrand:2014zea},
\mg~\cite{Alwall:2014hca}, or \sherpa~\cite{Sherpa:2019gpd}.  They
start from a fundamental Lagrangian and rely on perturbative quantum
field theory to provide the key part of a comprehensive simulation and
inference chain.  LHC simulations, including event generation, are a
challenging numerical task, so we expect improvements from modern
machine learning~\cite{Butter:2022rso,Plehn:2022ftl}. These
improvements cover precision and speed, which can at least in part be
interchanged, on track to meet the challenges by the upcoming HL-LHC
program.

Following the modular structure of event generators, we will use
modern neural networks to speed up expensive scattering amplitude
evaluations~\cite{Bishara:2019iwh,Badger:2020uow,Aylett-Bullock:2021hmo,Maitre:2021uaa,Winterhalder:2021ngy,Badger:2022hwf,Maitre:2023dqz}. Given
these amplitudes, we need to improve the phase-space integration and
sampling~\cite{Bendavid:2017zhk,Klimek:2018mza,Chen:2020nfb,Gao:2020vdv,Bothmann:2020ywa,Gao:2020zvv,Danziger:2021eeg,Heimel:2022wyj,Janssen:2023ahv,Bothmann:2023siu}.
Here, precise generative networks are extremely useful. Their
advantage in the LHC context is that they cover interpretable physics
phase spaces, for example scattering
events~\cite{Otten:2019hhl,Hashemi:2019fkn,DiSipio:2019imz,Butter:2019cae,Alanazi:2020klf,Butter:2021csz,Butter:2023fov},
parton
showers~\cite{deOliveira:2017pjk,Andreassen:2018apy,Bothmann:2018trh,Dohi:2020eda,Buhmann:2023pmh,Leigh:2023toe,Mikuni:2023dvk,Buhmann:2023zgc},
and detector
simulations~\cite{Paganini:2017hrr,deOliveira:2017rwa,Paganini:2017dwg,Erdmann:2018kuh,Erdmann:2018jxd,Belayneh:2019vyx,Buhmann:2020pmy,Buhmann:2021lxj,Krause:2021ilc,
  ATLAS:2021pzo,Krause:2021wez,Buhmann:2021caf,Chen:2021gdz,
  Mikuni:2022xry,ATLAS:2022jhk,Krause:2022jna,Cresswell:2022tof,Diefenbacher:2023vsw,
  Hashemi:2023ruu,Xu:2023xdc,Diefenbacher:2023prl,Buhmann:2023bwk,Buckley:2023rez,Diefenbacher:2023flw
}.

Generative networks for LHC physics can be trained on first-principle
simulations and are easy to ship, powerful in amplifying the training
data~\cite{Butter:2020qhk,Bieringer:2022cbs}, and --- most importantly
--- controlled and precise~\cite{Butter:2021csz,
  Winterhalder:2021ave,Nachman:2023clf,Leigh:2023zle,Das:2023ktd}.  Subsequent tasks
for these generative networks include event
subtraction~\cite{Butter:2019eyo}, event
unweighting~\cite{Verheyen:2020bjw,Backes:2020vka}, or
super-resolution enhancement~\cite{DiBello:2020bas,Baldi:2020hjm}.
Their conditional versions enable new analysis methods, like
probabilistic
unfolding~\cite{Datta:2018mwd,Bellagente:2019uyp,Andreassen:2019cjw,Bellagente:2020piv,Backes:2022vmn,Leigh:2022lpn,Raine:2023fko,Shmakov:2023kjj,Ackerschott:2023nax,Diefenbacher:2023wec},
inference~\cite{Bieringer:2020tnw,Butter:2022vkj,Heimel:2023mvw}, or
anomaly
detection~\cite{Nachman:2020lpy,Hallin:2021wme,Raine:2022hht,Hallin:2022eoq,Golling:2022nkl,Sengupta:2023xqy}.

Every LHC event generator covers phase space using a combination of
importance sampling and a multi-channel factorization.  To improve
phase-space sampling through modern machine learning, we need to
tackle both of these technical ingredients together. This observation
is the basis of the \madnis approach~\cite{Heimel:2022wyj}, which has independently been proposed by Refs.~\cite{Gao:2020vdv,Bothmann:2020ywa}.
Going beyond this proof of principle, we show in this paper how \madnis can
be implemented within \mg and by how much a classic
event generator setup can be improved. This is the logical next step after employing \madnis within the phase-space generator Chili~\cite{Bothmann:2023siu} and is similar to previous applications of neural importance sampling (NIS) in Sherpa~\cite{Gao:2020zvv}. In Sec.~\ref{sec:madnis} we
describe the updated \madnis framework, including an improved loss for
the network training in Sec~\ref{sec:madnis_loss}, a fast
initialization using classic importance sampling in
Sec.~\ref{sec:madnis_init}, and an efficient training strategy in
Sec.~\ref{sec:madnis_train}. In Sec.~\ref{sec:implement_feat} we then
provide detailed performance tests for a range of reference processes,
and always compared to an optimized \mg setup. In
Sec.~\ref{sec:implement_cw}, we show how the \madnis performance gains
can be understood in terms of amplitude patterns and symmetries, and
in Sec.~\ref{sec:implement_scale} we push our implementation to its
limits. As a spoiler, we will find that for challenging LHC processes,
\madnis can improve the unweighting efficiencies by an order of
magnitude, a performance metric which can be translated directly into
a speed gain. This gain is orthogonal to the development of faster
amplitude evaluations and hardware acceleration using highly
parallelized GPU farms~\cite{Valassi:2021ljk,Valassi:2022dkc,Bothmann:2023gew}.

\section{Improving \madnis}
\label{sec:madnis}

Before we discuss our improved ML-implementations, we briefly review
the basics of multi-channel Monte Carlo. We integrate a function $f
\sim \vert\mathcal{M}\vert^2$ over phase space
\begin{align}
    I[f] = \int \d^D x\,f(x)
    \qqqquad
    x \in \mathbb{R}^D \; .
    \label{eq:psinteg}
\end{align}
This integral can be decomposed introducing local channel weights
$\ai(x)$~\cite{Maltoni:2002qb,Mattelaer:2021xdr}
\begin{align}
  f(x)
  = \sum^{n_c}_{i=1} \ai(x) \; f(x)
  \qquad \mwith \quad \sum^{n_c}_{i=1} \ai(x) =1
  \quad \mand \quad \ai(x)\ge 0\;,
\label{eq:norm_alpha}
\end{align}
This \mg decomposition is different from the original multi-channel
method~\cite{Kleiss:1994qy, Weinzierl:2000wd}, which is used by
\sherpa~\cite{Sherpa:2019gpd}, or \whizard~\cite{Kilian:2007gr}, as
discussed in Ref.~\cite{Heimel:2022wyj}. The phase-space integral now
reads
\begin{align}
  I[f]
  = \sum^{n_c}_{i=1} I_i[f]
  = \sum^{n_c}_{i=1}\int \d^D x\;\ai(x)\,f(x) \; .
  \label{eq:multi-channel-mg1}
\end{align}
Next, we introduce a set of channel-dependent phase-space mappings
\begin{align}
x \in \mathbb{R}^D \quad 
\xleftrightarrow[\quad \leftarrow \gbar_i(z)\quad]{G_i(x)\rightarrow} 
\quad z\in[0,1]^D  \; ,
\label{eq:ps_mapping}
\end{align}
which parametrize properly normalized densities
\begin{align}
   g_i(x)=\left\vert\frac{\partial G_i(x)}{\partial x}\right\vert
   \qquad \mwith \quad 
   \int \d^D x\,g_i(x)=1  \; .
   \label{eq:channel_densities}
\end{align}
The phase-space integral now covers the $D$-dimensional unit cube and
can be sampled as
\begin{align}
\begin{split}
  I[f]
  &= \sum^{n_c}_{i=1}\int \left. \frac{\d^D z}{g_i(x)} \; \alpha_i(x) f(x) \right\vert_{x=\gbar_i(z)} \\
  &= \sum^{n_c}_{i=1}\int \d^D x \; g_i(x) \; \frac{\alpha_i(x) f(x)}{g_i(x)}
  \equiv \sum^{n_c}_{i=1} \left\langle \frac{\alpha_i(x) f(x)}{g_i(x)} \right\rangle_{x\sim g_i(x)}  \; .
\end{split}
  \label{eq:multi-channel-mg2}
\end{align}
Typically, physics-inspired channels are initialized with analytic mappings
and then refined by an adaptive algorithm, for instance
\vegas~\cite{Lepage:1977sw,Lepage:1980dq,Lepage:2020tgj, Ohl:1998jn,
  Brass:2018xbv}.  It exploits the fact that the
$I_i[f]$ are unchanged under the mappings $G_i$, but the variance
\begin{align}
\begin{split}
  \sigma_i^2\equiv\sigma_i^2\left[\frac{\alpha_i f}{g_i}\right] 
  = \left\langle \left( \frac{\alpha_i(x) f(x)}{g_i(x)}-I_i[f] \right)^2 \right\rangle_{x\sim g_i(x)}  \\
  = \left\langle  \frac{\alpha_i(x)^2 f(x)^2}{g_i(x)^2} \right\rangle_{x\sim g_i(x)} \; - I_i[f]^2
\end{split}
  \label{eq:def_sigma}
\end{align}
is minimized by the optimal mapping
\begin{align}
g_i(x)\big|_\text{opt}=\frac{\ai(x)f(x)}{I_i[f]}\; ,
\label{eq:optimal}
\end{align}
so the optimal form of $g_i(x)$ is only defined in relation to the
choice of $\alpha_i(x)$ and, consequently, $I_i[f]$.

We compute the Monte Carlo estimate of each integral using a discrete
set of points,
\begin{align}
  I_i[f]
  = \left\langle\frac{\alpha_i(x) f(x)}{g_i(x)} \right\rangle_{x\sim g_i(x)}
  \approx \frac{1}{N_i}\sum_{k=1}^{N_i}\left.\frac{\alpha_i(x_k) f(x_k)}{g_i(x_k)}\right\vert_{x_k\sim g_i(x)} \; .
\end{align}
In analogy, the variance gives us an estimate of the
uncertainty~\cite{Weinzierl:2000wd},
\begin{align}
  \Delta_{i,N_i}^2=\frac{\sigma_i^2}{N_i}
  &=\frac{1}{N_i}\left[
    \left\langle \frac{\alpha_i(x)^2 f(x)^2}{g_i(x)\,q_i(x)} \right\rangle_{x\sim q_i(x)}
      - \left\langle \frac{\alpha_i(x) f(x)}{q_i(x)} \right\rangle_{x\sim q_i(x)}^2 \right] \; .
  \label{eq:var_all}
\end{align}
Here, we rely on a different sampling density $q_i(x)\ne g_i(x)$. This
appears ad-hoc, but is needed to define a properly differentiable loss
function during training and will improve the
performance~\cite{Heimel:2022wyj}. When the variance is estimated from a finite sample, a correction factor $N_i/(N_i-1)$
has to be introduced to ensure an unbiased estimator.  As the complete
integral $I[f]$ is the sum of statistically independent $I_i[f]$, its
uncertainty and variance are
\begin{align}
  \Delta^2_{N} &= \sum^{n_c}_{i=1} \Delta^2_{i,N_i}=\sum^{n_c}_{i=1} \frac{\sigma^2_i}{N_i}
  \qquad \mand \qquad 
  \sigma_\text{tot}^2
  = N\Delta^2_{N}
  = \sum^{n_c}_{i=1} \frac{N}{N_i}\sigma^2_i\; .
  \label{eq:var_channel}
\end{align}
%

\subsection{ML implementation}
\label{sec:madnis_ml}

Starting from Eq.\eqref{eq:multi-channel-mg2}, \madnis encodes the
multi-channel weight $\ai(x)$ and the channel mappings $G_i(x)$ in
neural networks.

\subsubsection*{Neural channel weights}

First, \madnis employs a channel-weight neural network (CWnet) to
encode the local multi-channel weights defined in
Eq.\eqref{eq:norm_alpha},
\begin{align}
   \ai(x)\equiv\aix(x) \; ,
  \label{eq:trained_weights}
\end{align}
where $\xi$ denotes the network parameters. The best way to implement
the normalization condition from Eq.\eqref{eq:norm_alpha} in the
architecture is~\cite{Heimel:2022wyj}
\begin{align}
  \aix(x) =
  \softmax A_{i\xi}(x)
  \equiv
  \frac{\exp A_{i\xi}(x)}{\sum_j \exp A_{j\xi}(x)} \; ,
\end{align}
where $A_i(x)$ is the unnormalized network output.
As the channel weights vary strongly over phase space, it helps the
performance to learn them as a correction to a physically motivated
prior assumption. Two well-motivated and normalized priors in \mg are
\begin{align}
\begin{split}
    \ai^{\text{MG}}(x)&= \frac{|\mathcal{M}_i(x)|^2}{\sum_j |\mathcal{M}_j(x)|^2} \\
    \ai^{\text{MG}}(x) &= \frac{P_i(x)}{\sum_j P_j(x)} 
  \quad \text{with} \quad
  P_i(x) =  \prod_{k \in \text{prop}} \frac{1}{|p_k(x)^2-m_k^2 -\imag m_k\Gamma_k|^2} \; .
\end{split}
\label{eq:mg_prior}
\end{align}
They define the single-diagram enhanced multi-channel
method~\cite{Maltoni:2002qb,Mattelaer:2021xdr}.  Relative to either of
them we only learn a correction to the prior, \ie
\begin{align}
\begin{split}
  \aix (x) &= \softmax\!\left[\log \alpha^{\text{MG}}_i(x)+ A_{i\xi}(x)\right] = \frac{\alpha^{\text{MG}}_i(x) \; \exp A_{i\xi}(x)}{\sum_j [ \alpha^{\text{MG}}_j(x) \; \exp A_{j\xi}(x)]} \; .
\end{split}
\label{eq:delta_channel}
\end{align}
We initialize the network parameters $\xi_0$ such that $A_{i\xi_0}(x)=0$.

\subsubsection*{Neural importance sampling}

Second, we combine analytic channel mappings and a normalizing flow
for a mapping from a latent space $z$ to the phase space $x$,

\begin{align}
x \in \mathbb{R}^D 
\xleftrightarrow[\hspace{1.1cm}]{\text{analytic}} 
y\in[0,1]^D
\xleftrightarrow[\hspace{1.1cm}]{\text{INN}}
z\in[0,1]^D \; .
\label{eq:def_inn}
\end{align}
This chain replaces \vegas with an invertible neural network (INN)~\cite{inn}, an incarnation of a normalizing flow equally fast in both directions. This INN is used for latent space refinement,
where the INN links two unit hypercubes $[0,1]^D$.  \madnis improves
the physics-inspired phase-space mappings by training an INN as part of
\begin{align}
  z = \Git(x) \qquad \text{or} \qquad x = \gbar_{i \theta}(z) \; ,
  \label{eq:trained_mappings}
\end{align} 
with the network weights $\theta$.  Similar to the multi-channel
weights, we use physics knowledge to simplify the task of the
normalizing flow and improve its efficiency.

A crucial condition for using a normalizing flow for phase-space
integration is its equally fast evaluation in both
directions~\cite{Heimel:2022wyj}. The building blocks of the
best-suited flows are coupling layers.  The INN task in
Eq.\eqref{eq:def_inn} can, in principle, be realized with any
invertible coupling layer and an additional sigmoid layer. Coupling
layers~\cite{coupling1,coupling2,inn} split the input $y$ into two
parts, $y=(y^A,y^B)$, and define the forward and inverse passes as
\begin{align}
    \begin{pmatrix}
    z^A_r \\ z^B_s
    \end{pmatrix}
    =
    \begin{pmatrix}
    y^A_r \\ C(y^B_s; u_{s \theta}(y^A))
    \end{pmatrix} 
    \qquad \Leftrightarrow \qquad
    \begin{pmatrix}
    y^A_r \\ y^B_s 
    \end{pmatrix}
    =
    \begin{pmatrix}
    z^A_r \\ C^{-1}(z^B_s; u_{s \theta}(z^A))
    \end{pmatrix} \; ,
    \label{eq:rqs_block}
\end{align}
The component-wise coupling transform $C$ acts on the learned function
$u_\theta$, is invertible and has a tractable Jacobian. The Jacobian
of the full mapping is
\begin{align}
  g(y)
  =\prod_{s} \frac{\partial C(y^B_s; u_{s \theta}(y^A))}{\partial y^B_s} \;.
\end{align}
For a stable numerical performance we use rational-quadratic spline
(RQS) blocks~\cite{durkan2019neural}, where each bin is a
monotonically-increasing rational-quadratic (RQ) function. They
provide superior expressivity and are naturally defined on a compact
interval $[a,b]$.  We split the unit interval $[0,1]$ in $K$ bins
with $K+1$ boundaries or knots $(y_s^{(k)},z_s^{(k)})$
$(k=0,\dots,K)$, with fixed endpoints $(0,0)$ and $(1,1)$. The $K$
widths $w_s$ and heights $h_s$ of the bins and the $K+1$ derivatives
$d_s$ at the boundaries are constructed from the output of the
network,
\begin{align}
 u_{s \theta}(x^A) = (\Theta^w_s,\Theta^h_s,\Theta^d_s) \; .
\end{align}
They are normalized as
\begin{align}
\begin{split}
    w_s &= \softmax \Theta_s^w  \\
    h_s &= \softmax \Theta_s^h  \\
    d_s &=\frac{\softplus \Theta_s^d}{\log 2}\equiv\frac{\log\left(1+\exp \Theta_s^d\right)}{\log 2} \; .
\end{split}
\label{eq:rqs_set_bins}
\end{align}
In contrast to the original Ref.~\cite{durkan2019neural}, we add
learnable derivatives at the boundaries to increase expressivity, and
introduce the normalization of $d_s$ such that $\Theta_s^d=0$ is
associated with a unit derivative $d_s=1$. The $w$, $h$ and $d$ are
then used to construct the RQ function $C$ and its
derivative~\cite{durkan2019neural}.

In Eq.\eqref{eq:rqs_block} we see that the coupling layer describes
each dimension $z_s^B$ using a learned function of all $y^A=z^A$. This
way, it cannot describe correlations between different directions
$z^B_s$. To encode correlations between all dimensions, we stack
multiple coupling layers and permute the elements between them. The
permutation changes which components are combined within $y^A$ and
$y^B$. We use a deterministic set of permutations based on a
logarithmic decomposition of the integral dimension~\cite{Gao:2020vdv,
  Heimel:2022wyj}, which ensures that any element is conditioned on
any other element at least once. The number of coupling blocks then
scales with $\log D$.

\subsection{Multi-channel loss}
\label{sec:madnis_loss}

After encoding channel weights and importance sampling as neural
networks,
\begin{align}
\ai(x) \equiv \aix(x)
\qquad \mand \qquad 
g_i(x)\equiv \git(x) \; ,
\end{align}
the integral in Eq.\eqref{eq:multi-channel-mg2} is estimated as
\begin{align}
  I[f]
  = \sum^{n_c}_{i=1}\left\langle\frac{\aix(x) f(x)}{\git(x)}\right\rangle_{x\sim \git(x)} \; .
\end{align}
The common optimization task for both networks is given by
Eq.\eqref{eq:optimal}.  The relative contribution per sub-integral
changes when we train the two networks, ruling out the use of any
$f$-divergence~\cite{Nielsen:2014} when optimizing the channel weights $\aix(x)$.

As the optimization is performed on batches with size $b\ll N$, we use
a loss function that does not scale with $b$ and is independent of the
number of sampled points. Moreover, the sampling density in the loss
function $x\sim q_i(x)$ cannot coincide with the learned
mapping~\cite{Heimel:2022wyj}.  We can directly minimize the variance
as the loss function
\begin{align}
\begin{split}
\loss_\text{variance}
&=\sum^{n_c}_{i=1} \frac{N}{N_i}\sigma^2_i \\
&=\sum^{n_c}_{i=1}\frac{N}{N_i}\left(
    \left\langle \frac{\aix(x)^2 f(x)^2}{\git(x)\,q_i(x)} \right\rangle_{x\sim q_i(x)}
      - \left\langle \frac{\aix(x) f(x)}{q_i(x)} \right\rangle_{x\sim q_i(x)}^2 \right) \;.
\end{split}
      \label{eq:var_loss}
\end{align}
In practice, $q_i(x)\simeq \git(x)$ allows us to compute the loss as
precisely as possible and stabilizes the combined
online~\cite{Butter:2022lkf} and buffered
training~\cite{Heimel:2022wyj}.

A critical hyperparameter in the variance loss is the distribution of
sample points, $N_i$, during training and integral evaluation.  Its
optimal choice depends on the $\sigma^2_i$, and with that on $\ai$ and
$g_i$. While the latter are only known numerically, the optimal choice
of $N_i$ can be computed by minimizing the loss with respect to $N_i$,
given $N=\sum_i N_i$~\cite{Press:1989vk,Weinzierl:2000wd}
\begin{align}
  N_i
 = N \frac{\sigma_{i}}{\sum_j \sigma_{j}} 
 \qqquad \Leftrightarrow \qqquad 
 \frac{N}{N_i} = \frac{\sum_j \sigma_{j}}{\sigma_{i}}  \; .
 \label{eq:stratified-sampling}
\end{align}
This known result from stratified sampling defines the improved
\madnis loss
\begin{align}
\begin{split}
    \loss_\madnis 
    &= \sum_{i=1}^{n_c} \left(\sum_{j=1}^{n_c}\sigma_j\right)\sigma_i= \left[\sum_{i=1}^{n_c} \sigma_i\right]^2 \\
    &= \left[\sum_{i=1}^{n_c} \left(
    \left\langle \frac{\aix(x)^2 f(x)^2}{\git(x)\,q_i(x)} \right\rangle_{x\sim q_i(x)}
      - \left\langle \frac{\aix(x) f(x)}{q_i(x)} \right\rangle_{x\sim q_i(x)}^2 \right)^{1/2} \right]^2 
    \;.
\end{split}
    \label{eq:madnis_loss}
\end{align}
The physics information used to construct a set of integration
channels is usually extracted from Feynman diagrams.  \mg
automatically groups Feynman diagrams or channels, which are linked by
permutations of the final-state particles. Because of the
complications through the parton densities it does not consider
permutations in the initial state.  Following the \mg approach,
symmetry-related \madnis channels share the same phase space mapping,
but the channel-weight network has separate outputs for each
channel. The $\sigma_i$ in the loss can be defined for individual
channels or for channel groups. For processes with many channels we
find that using sets of channels stabilizes the training, so we resort
to this assignment as our default.

\subsection{\vegas initialization}
\label{sec:madnis_init}

\vegas~\cite{Lepage:1977sw,Lepage:1980dq,Lepage:2020tgj} is the
classic algorithm for importance sampling to compute high-dimensional
integrals.  For approximately factorizing integrands \vegas is
extremely efficient and converges much faster than factorized neural
importance sampling, so we use use a \vegas pre-training to initialize
our normalizing flow.

\begin{figure}[t]
    \includegraphics[width=0.49\textwidth]{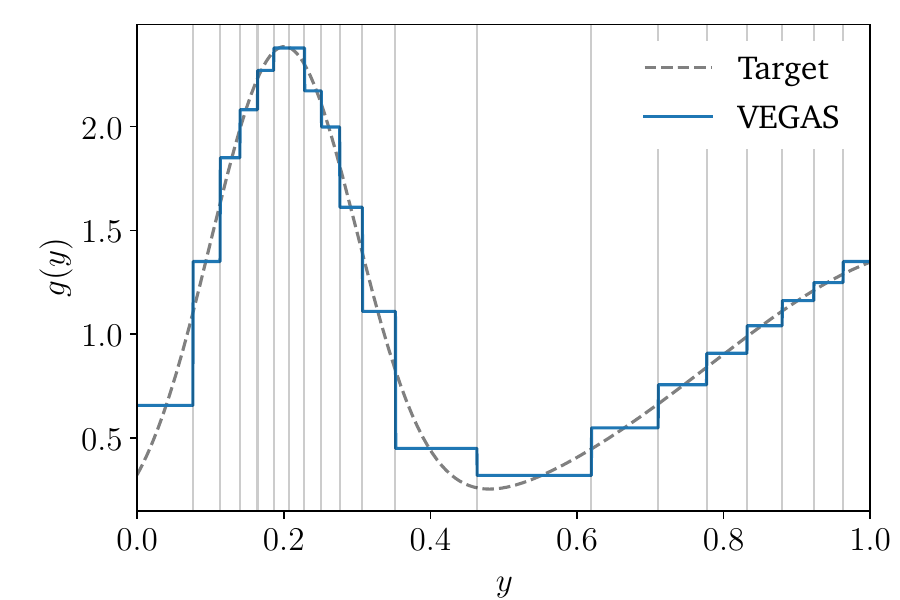}
    \includegraphics[width=0.49\textwidth]{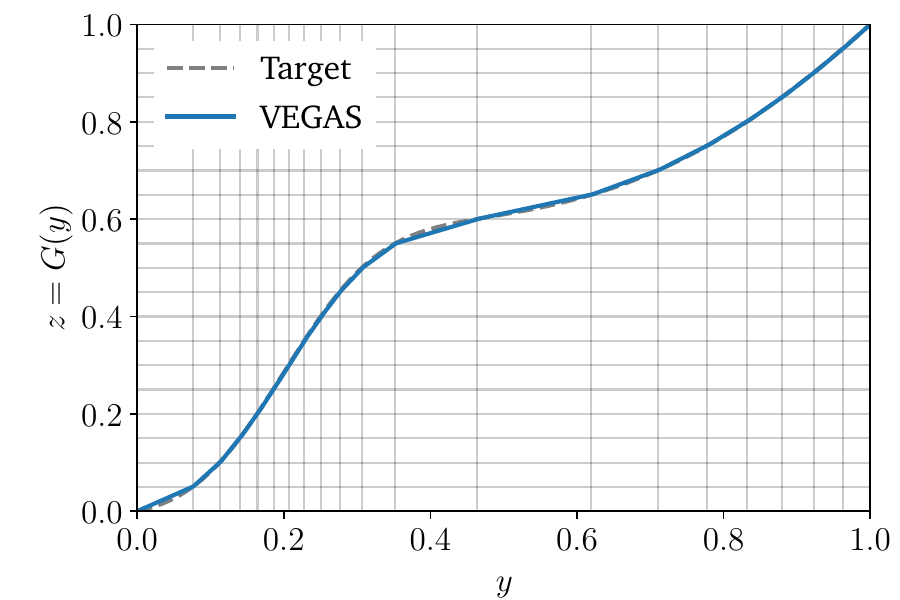}\\
    \includegraphics[width=0.49\textwidth]{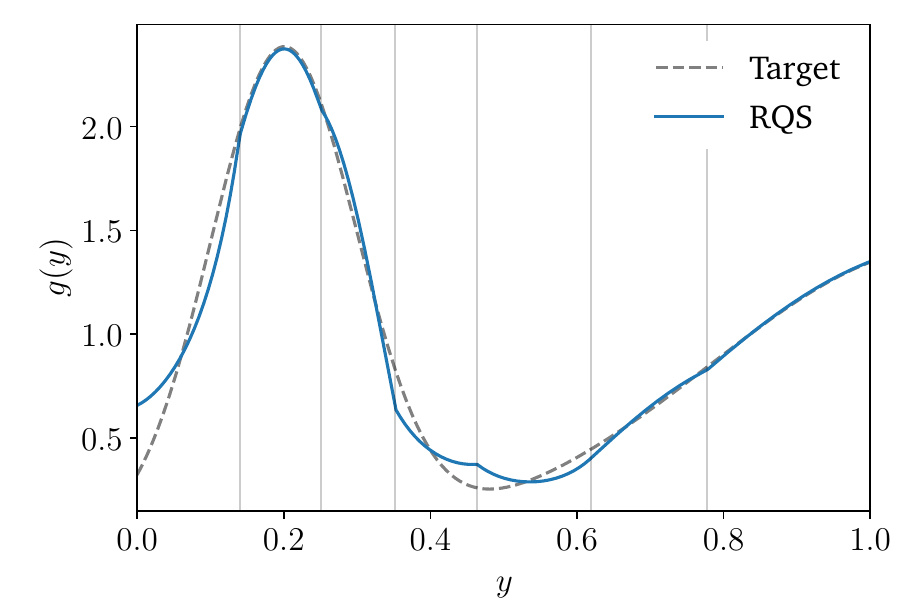}
    \includegraphics[width=0.49\textwidth]{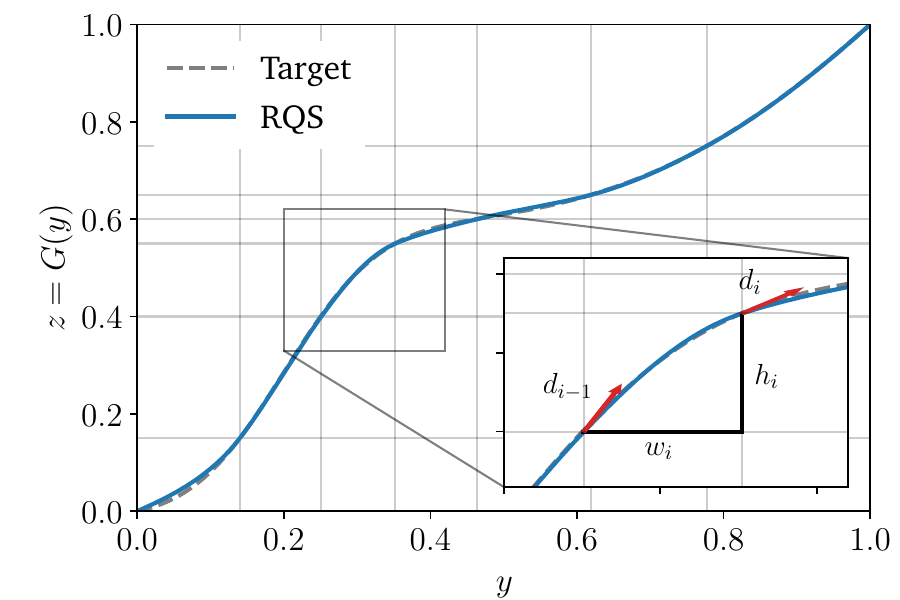}
    \caption{Top: learned \vegas density $g(y)$ with 20 bins (left)
      and its transformation $G(y)$ (right). Bottom: the RQS density
      $g(y)$ after embedding the \vegas grid and performing bin
      reduction to 7 bins (left). The right plot shows the
      corresponding mapping $G(y)$ including a zoom-in box
      illustrating the definition of the widths $w_i$, heights $h_i$
      of the bins and the derivatives $d_i$ on the bin
      edges.}
    \label{fig:vegas_grid}
\end{figure}

The standard \vegas implementation integrates the unit cube, $z_s \in
[0,1]$ and relies on an assumed factorization of the phase space
dimensions, made explicit for the phase space mapping and the
corresponding sampling distribution in the second step of
Eq.\eqref{eq:def_inn}
\begin{align}
    g(y) = \prod_{s=1}^D g(y_s) \; .
    \label{eq:vegas_factorize}
\end{align}
In analogy to Eq.\eqref{eq:ps_mapping} it encodes sampling
distributions $g$ for each dimension by dividing the unit interval
into $K$ bins of equal probability but different widths $w_k$ with
$\sum_k w_k=1$. In each interval,
\begin{align}
  y \in [W_{k-1}, W_k]
  \qquad \mwith \qquad 
  W_k = \sum_{m=1}^k w_m \; ,
\end{align}
\vegas samples uniformly,
\begin{align}
  g(y) = \frac{1}{K w_k}
  \qquad \mfor \qquad
  y \in [W_{k-1}, W_k] \; ,
\end{align}
such that each bin integrates to $1/K$, and their sum from
Eq.\eqref{eq:channel_densities} to
\begin{align}
  \int_0^1 \d y\; g(y) = \sum^K_{k=1} w_k \; \frac{1}{K w_k} = 1 \; .
\end{align}
In one dimension, the integrated mapping $G(x)$ from
Eq.\eqref{eq:ps_mapping} is a piece-wise linear function, or linear
spline,
\begin{align}
  G(y) = (k-1) \; \frac{1}{K} + \frac{y - W_{k-1}}{K w_k}
  \qquad \mfor \qquad 
  y \in [W_{k-1}, W_k] \; .
\end{align}
The \vegas algorithm iteratively adapts the bin widths $w_k$ such that
the distribution $g(y)$ matches the integrand.  We illustrate $g$ and
$G$ with $K=20$ bins for a Gaussian mixture model in the upper panels
of Fig.~\ref{fig:vegas_grid}.

If we want to use \vegas to pre-train an INN-mapping of two unit
cubes, we need to relate the \vegas grid to the RQ splines introduced
in Sec.~\ref{sec:madnis_ml}. Following Eq.\eqref{eq:rqs_set_bins},
we need to extract the bin widths $w$, the bin heights $h$, and the
derivatives on the bin edges $d$ from a \vegas grid.  The width in $y$
are the same as the \vegas bin sizes, the heights are equal for all
bins, and the derivatives can be estimated as the ratio of the bin
heights and widths,
\begin{alignat}{3}
    \text{widths:} && \qquad
    w_k &&& \qquad \mfor \quad k=1,\dots,K  \notag \\
    \text{heights:} && \qquad
    h_k &= \frac{1}{K} && \qquad \mfor \quad k=1,\dots,K  \notag \\
    \text{endpoints:} && \qquad
    d_0 &= \frac{h_1}{w_1} = \frac{1}{K w_1} \qquad 
    d_{K} = \frac{h_{K}}{w_{K}} = \frac{1}{K w_K} && \notag \\
    \text{internal points:} && \qquad
    d_k &= \frac{h_{k+1} + h_k}{w_{k+1} + w_k} = \frac{2}{K(w_{k+1} + w_k)} 
    &&\qquad \mfor \quad k=1,\dots,K-1\; .
\end{alignat}
Because of the higher expressivity of the RQ splines, they need much
fewer bins than \vegas. For instance, all bins in \vegas have equal
probability, meaning that more bins than necessary are used in regions
of high probability but with little change in the derivative of the
transformation. We therefore reduce the number of bins by repeatedly
merging adjacent bins until the desired number of target bins is
reached:
\begin{enumerate}
    \item Calculate the absolute difference between the slopes of adjacent bins,
        \begin{align}
            \Delta_k = \left| \frac{w_k}{h_k} - \frac{w_{k+1}}{h_{k+1}} \right| \; .
        \end{align}
    \item Choose the lowest $\Delta_k$, \ie the two bins $k$ and $k+1$
      with the most similar average slope. We introduce an additional cutoff,
      to prevent very large bins, unless all smaller bins are already
      merged.
    \item Reduce the number of bins by one
    \begin{align}
    \begin{split}
        w &\leftarrow (w_1, \ldots, w_{k-1}, w_k + w_{k+1}, w_{k+2}, \ldots, w_{K})\;, \\
        h &\leftarrow (h_1, \ldots, h_{k-1}, h_k + h_{k+1}, h_{k+2}, \ldots, h_{K})\;, \\
        d &\leftarrow (d_0, \ldots, d_{k-1}, d_{k+1}, \ldots, d_{K})\;, \\
        K &\leftarrow K-1 \;.
    \end{split}
    \end{align}
\end{enumerate}
The lower panels of Fig.~\ref{fig:vegas_grid} show $g$ and $G$ for the
RQ spline after applying the bin reduction algorithm to reduce the
number of bins to $K=7$. We show $w$, $h$ and $d$ for one bin.

As shown in Eq.\eqref{eq:rqs_block}, a RQS spline block includes a
trainable sub-network. It encodes the bin widths $w$, heights $h$ and
derivatives $d$. By inverting the normalization from
Eq.\eqref{eq:rqs_set_bins}, setting the weight matrix of the final
sub-network layer to zero and assigning our results for $\Theta^w$,
$\Theta^h$ and $\Theta^d$ to the bias vector, we can initialize the
spline block to behave like the \vegas grid. This has to be done for
the final two coupling blocks (in the sampling direction), one for
each half of the dimensions.

\subsection{Training strategies}
\label{sec:madnis_train}

\subsubsection*{Online \& buffered training}

\madnis supports two training modes~\cite{Heimel:2022wyj}. First, in
the online training mode generated events are immediately used to
update the network weights. Their integrands, channel weights, and
Jacobians are stored for later use in the second, buffered training
mode. For that, only the weight update is performed. Online
training is typically more stable, especially in the early stages,
while buffered training is much faster in cases where the integrand is
computationally expensive.  We alternate between online and buffered
training and keep the buffer at a fixed size, so the oldest sample
points are replaced through online training. The impact of the
buffering is given by $R_@$, the ratio of the number of weight updates
from online training to all weight updates. For expensive integrands,
$R_@$ determines the speed gain relative to online training only.
For the results shown in
Sec.~\ref{sec:implement} we use a buffer size of 1000 batches and
start with online training until the buffer is filled. We then train
alternately on the full buffer and online, where we aim to replace
either a quarter or half of the buffer, corresponding to gain factors
$R_@ = 5$ and $R_@ = 3$.

Learning and correcting the channel weights changes the normalization
for each channel.  Since this normalization is a function of all
channel weights, the complete set has to be stored as part of the
buffered training. Storing $\mathcal{O}(1M)$ phase space points for a
process with $\mathcal{O}(1k)$ channels requires several gigabytes, so
this becomes prohibitive. On the other hand, we typically find that
for a given sampling only a few channels contribute
significantly. Instead of storing all weights we then store the
indices and weights $(i,\alpha_i)$ of the $m<n_c$ channels with the
largest weights. The channel that was used to generate the sample
always has to be included in this list, even if it is not among the
$m$ largest weights. To optimize the network on such a buffered
sample, we set the other $\ai$ to zero and adjust the normalization
accordingly. Because this approximation is only used during buffered
training epochs, but not during online training, integration or
sampling, it does not introduce a bias.

\subsubsection*{Stratified training \& channel dropping}

The variance of a channel scales with its cross section, and even
if different channels contribute similarly to the total cross section, their
variances can still be very imbalanced. The variance loss will
be dominated by channels with a large variance, leading to large and noisy training
gradients if the number of points in these channels is not sufficiently large. We improve our training by adjusting the number of points
per channel and allow for channels to be dropped altogether.

Because the variance of the channels changes during the training of
the channel weights and during the training of the importance
sampling, we track running means of the channel variances and use them
to adjust the number of points per channel during online training. To
ensure a stable training we first distribute a fraction $r$ of points
evenly among channels. The remaining part of the sample is distributed
according to the channel variances, following the stratified sampling
defined in Eq.\eqref{eq:stratified-sampling}.  This means $r$
interpolates between uniform sampling ($r=1$) and stratified sampling
($r=0$). Note that in the presence of phase space cuts we cannot be
sure that there will be points with non-vanishing weights in a given
channel. For the results shown in Sec.~\ref{sec:implement} we ensure a
stable estimate of the channel variance by first training with uniform
$N_i$ for 1000 batches and then compute the variance with a running
mean over the last 1000 batches.

We also go a step further and drop channels with a very small
contribution to the cross section. For this, we track the running mean
over the $I_i$ per channel and allow for channels to be dropped after
every training epoch. We define a fraction of the total cross section,
typically $10^{-3} \times I$, which can be neglected by dropping the
corresponding channels. Next we sort all channels by $I_i$ and drop
channels, starting from the smallest contributions, until this
fraction is reached.  If we drop a channel we also remove its
contribution to the buffered sample and adjust the normalization of
the channel weights. This way no training time is invested into
dropped channels and the result remains unbiased.

\section{Implementation and benchmarks}
\label{sec:implement}

We call \mg from \madnis to compute the
matrix element and parton densities, the initial phase space mapping,
and the initial channel weights used in Eq.\eqref{eq:delta_channel}.
For each event the inputs to this \mg call are a vector of random
numbers $r_s$ with $s = 1 \dots D$ and the index $i$ of the channel
used to sample the event.  The \mg output is the concatenation of all
4-momenta $p$, the event weight $w$, and the set of channel weights
$\alpha_j^\text{MG}$ with $j = 1 \dots n_c$.  
All \madnis hyperparameters used for our benchmark study are given in
the Appendix, Tab.~\ref{tab:hyperparams}. We note, while trained on significantly less training points than the flow, \vegas reaches its optimum faster due to its simpler optimization and construction. We further emphasize, that \vegas is trained much longer and on more points than usually done in \mg, guaranteeing it has reached its optimum. In detail, as indicated in Tab.~\ref{tab:hyperparams}, we train \vegas for 7 iterations (default is 3) using 50k samples per iteration. We verified that increasing this number by a factor of 10 did not further improve the performance of \vegas.  Currently, our interface is not
fully optimized and is still a speed bottleneck. Furthermore, our
setup does not support multiple partonic processes yet. Hence, we will
not show run time comparisons in this work and will use processes with
a fixed partonic initial state.

\subsection{Reference processes}
\label{sec:implement_proc}

To benchmark the \madnis performance we choose a set of realistic and
challenging hard processes,
\begin{alignat}{4}
  \text{Triple-W}&  \qquad&  
\Pu \Pdbar &\to \PWp \PWp \PWm \quad &&&&
\notag \\
\text{VBS}&  \qquad &  
\Pu \Pc &\to \PWp \PWp\;\Pd\Ps \quad &&&&
\notag \\
\text{W+jets}&  \qquad  &
\Pg \Pg &\to \PWp \Pd \bar{\Pu} \qquad &
\Pg \Pg &\to \PWp \Pd \bar{\Pu}\Pg \qquad &
\Pg \Pg &\to \PWp \Pd \bar{\Pu}\Pg\Pg
\notag \\
\text{$\Pt\Ptbar$+jets}&  \qquad  & 
\Pg \Pg &\to \Pt\Ptbar+\Pg \qquad &
\Pg \Pg &\to \Pt\Ptbar+\Pg\Pg \qquad &
\Pg \Pg &\to \Pt\Ptbar+\Pg\Pg\Pg\; ,
\label{eq:proc}  
\end{alignat}
The produced heavy particles are assumed to be stable. The motivation
for VBS and Triple-W production is that they are key to understand
electroweak symmetry breaking, and that their large number of
gauge-related Feynman diagrams will challenge our framework with
potentially large interference. Next, we choose W+jets and
$\Pt\Ptbar$+jets production to study the scaling with additional
jets. In Tab.~\ref{tab:processes} we show the number of Feynman
diagrams and the number of \mg channels for all processes. For
instance, \mg does not construct a separate channel for four-point
vertices, so the number of channels is smaller than the number of
diagrams. Channels which only differ by a permutation of the
final-state momenta are combined into groups, further reducing the
numbers of mappings. Finally, we show the number of channels that
remain active after a \madnis training, as discussed in
Sec.~\ref{sec:implement_cw}.

\begin{table}[t]
    \centering
    \begin{small} \begin{tabular}[t]{ll|rrrr}
    \toprule
    Process & & \# diagrams & \# channels & \# channel groups
    & \# active channels\\
    \midrule
    Triple-W & $\Pu \Pdbar \to \PWp \PWp \PWm$ & 17 & 16 & 8 & \cbox{2 $\ldots$ 4}\\
    VBS & $\Pu \Pc \to \PWp \PWp\;\Pd\Ps$ & 51 & 30 & 15 & \cbox{4 $\ldots$ 6}\\
    W+jets & $\Pg \Pg \to \PWp\, \Pd \bar{\Pu}$ & 8 & 8 & 4 & \cbox{6}\\
    & $\Pg \Pg \to \PWp\, \Pd \bar{\Pu}\Pg$ & 50 & 48 & 24 & \cbox{12 $\ldots$ 16}\\
    & $\Pg \Pg \to \PWp\, \Pd \bar{\Pu}\Pg\Pg$ & 428 & 384 & 108 & \cbox{28 $\ldots$ 51}\\
    $\Pt\Ptbar$+jets & $\Pg \Pg \to \Pt\Ptbar+\Pg$ & 16 & 15 & 9 & \cbox{4 $\ldots$ 6}\\
    & $\Pg \Pg \to \Pt\Ptbar+\Pg\Pg$ & 123 & 105 & 35 & \cbox{12}\\
    & $\Pg \Pg \to \Pt\Ptbar+\Pg\Pg\Pg$ & 1240 & 945 & 119 & \cbox{60 $\ldots$ 72}\\
    \bottomrule
    \end{tabular} \end{small}
    \caption{Number of Feynman diagrams, channels and channel groups
      after accounting for symmetries. The last column shows the
      number of channels that remain active after \madnis channel
      dropping. Its range reflects ten independent trainings.}
    \label{tab:processes}
\end{table}

\subsection{Benchmarking \madnis features}
\label{sec:implement_feat}

To see the gain from each of the \madnis features introduced in
Sec.~\ref{sec:madnis} we apply them to our reference processes one by
one. Our baseline is \vegas, combined with \mg channel mappings and
channel weights. Again, for a fair comparison, we also optimize \vegas to the
best of knowledge and work with more iterations and larger samples
than the \mg default. The \vegas optimization is still
significantly faster than \madnis, but our goal is to provide a
pre-trained \madnis generator and the training time is amortized for generating large numbers of events, so the generation time is our only
criterion.

We use two metrics for our comparison: (i) the relative standard
deviation $\sigma / I$ minimized by stratified sampling, as is it
independent from the sampling statistics and cross section; (ii) the
unweighting efficiency $\epsilon$~\cite{Gao:2020zvv}, where the
maximum weight is determined by bootstrapping and taking the
median. For each, we use ten independent trainings and compute the
means and standard deviations as an estimate of the stability of the
training. We note that all obtained results, both integrated as well as differential cross-sections, agree with the default \mg output and yield the correct result.

\begin{figure}[t]
    \includegraphics[width=0.495\textwidth]{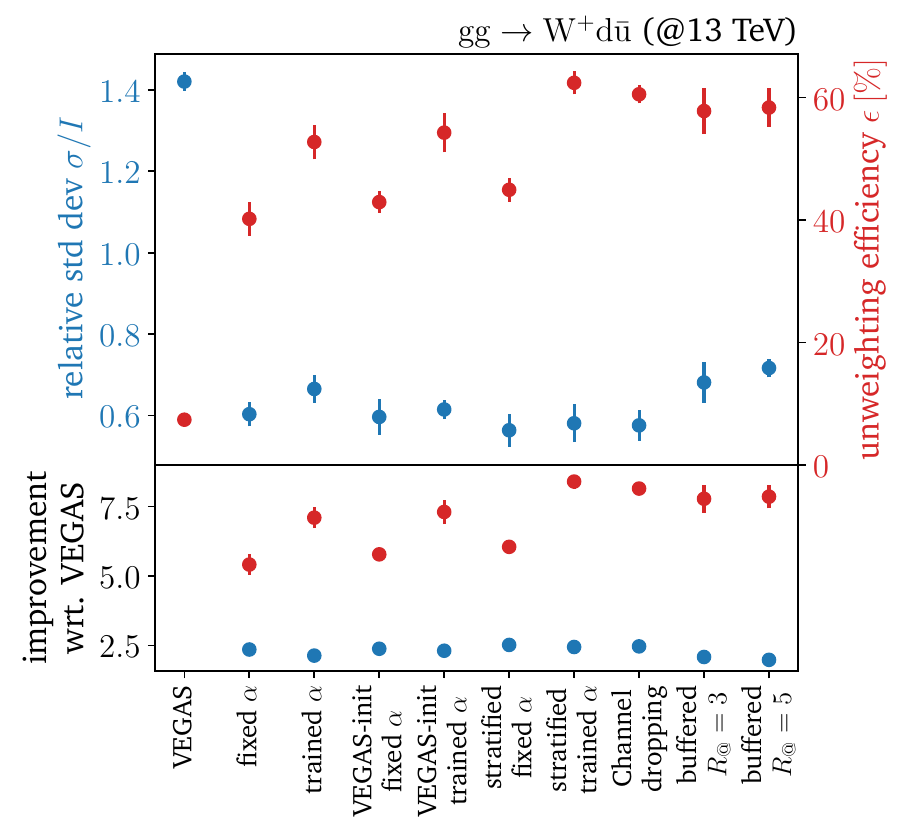}
    \includegraphics[width=0.495\textwidth]{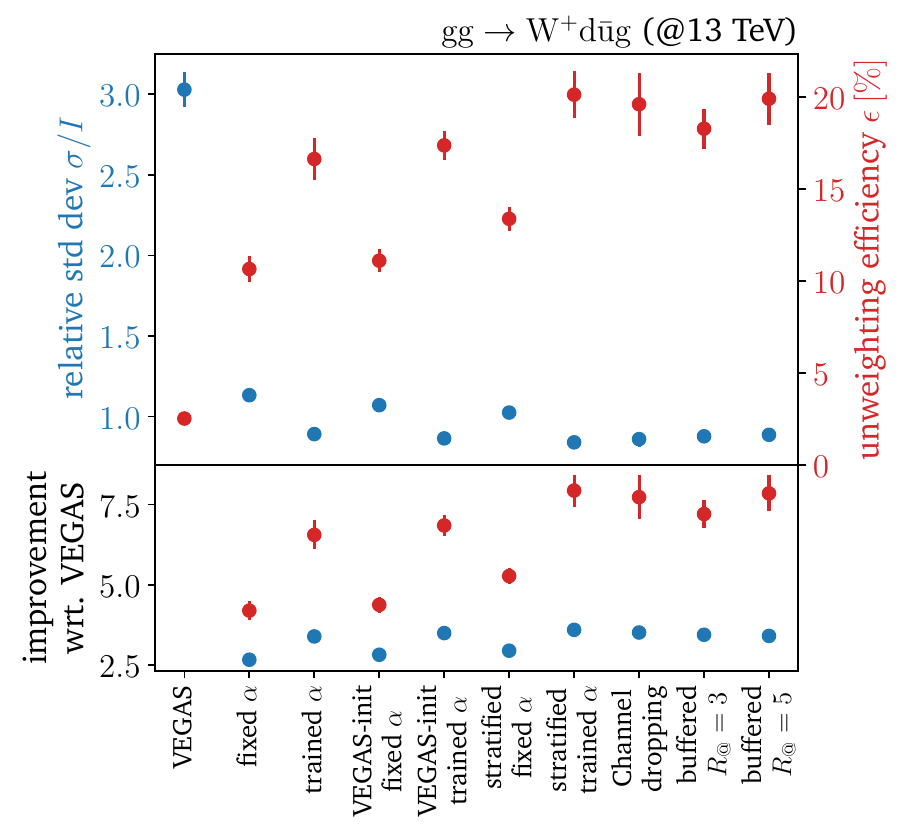}\\
    \includegraphics[width=0.495\textwidth]{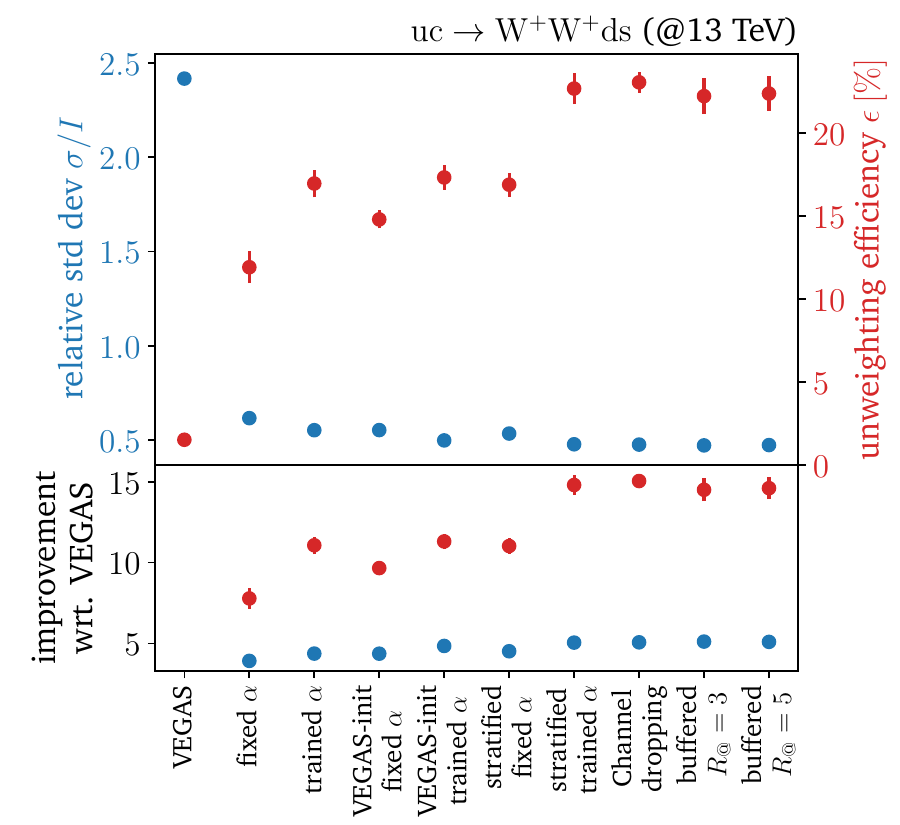}
    \includegraphics[width=0.495\textwidth]{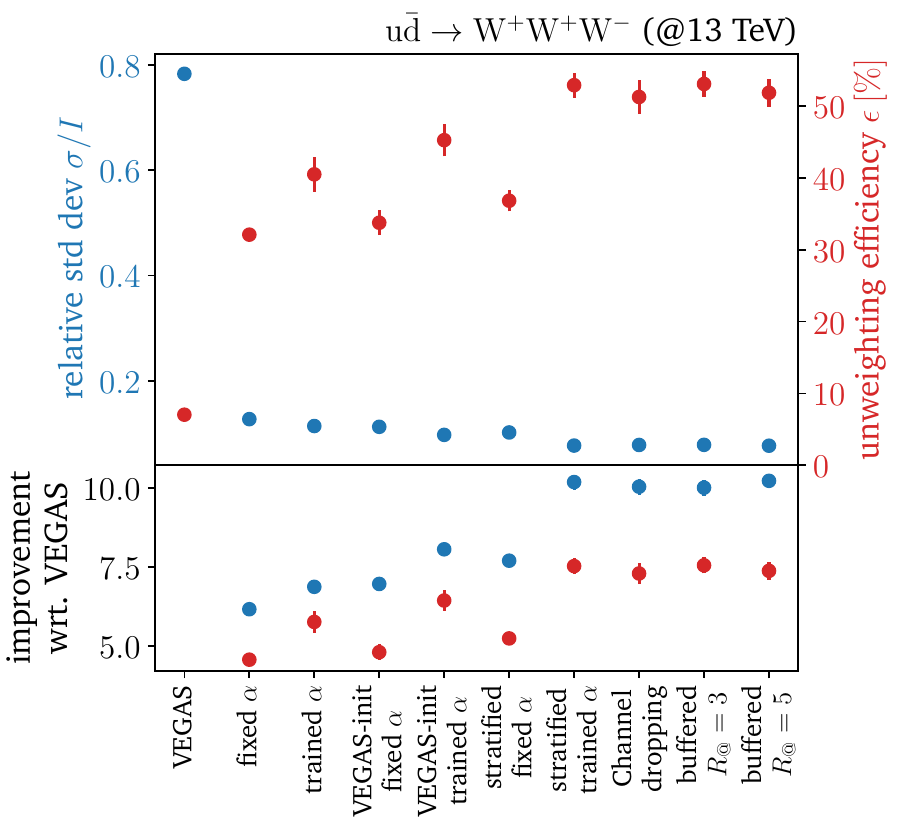}\\
    \caption{Relative standard deviation and unweighting efficiency
      for W+2jets, W+3jets, VBS and Triple-W for various combinations
      of \madnis features.}
    \label{fig:process_comparison}
\end{figure}

In Fig.~\ref{fig:process_comparison} we show the results of our
benchmarking for the four different processes with gauge bosons. For
the more time-consuming top pair processes without any specific
challenges we will provide the final improvements below. 

First, we show the results from a simple setup where only the channel
mappings are trained, but the channel weights are kept fixed,
Sec.~\ref{sec:madnis_ml}. Already here we see a sizeable gain over
the standard method.  Next, we also train the channel weights, and
observe a further significant gain. For instance, the unweighting
efficiency for VBS now reaches $20\%$, up from a few per-cent from the
standard method and by more than a factor ten.

Next, we stick to the trainings without and with adaptive channel
weights, but combine them with the \vegas-initialization from
Sec.~\ref{sec:madnis_init}. For all processes, the gain compared to
the standard \vegas method stabilizes, albeit without major
improvements. This changes when we include stratified training, as
introduced in Sec.~\ref{sec:madnis_train}. For all processes,
stratified sampling with trained channel weights lead to another
significant performance gain, up to a factor 15 for the VBS process.

Finally, we add channel dropping and buffered training, to reduce the training time. For
time-wise gain factors $R_@ = 3$ and $R_@ = 5$ the improvement of the
full \madnis setup over the standard \vegas and \mg integration
remains stable. For processes with a large number of channels,
channel dropping leads to a more stable training, while providing an equally good performance for the processes shown in Fig.~\ref{fig:process_comparison}. The fact that
the performance gain for the unweighting efficiency is much larger
than for the relative standard deviation reflects the sensitivity of
the unweighting efficiency to the far tails of the weight
distribution.

\subsection{Learning from channel weights}
\label{sec:implement_cw}

\begin{figure}[t]
    \includegraphics[width=0.495\textwidth]{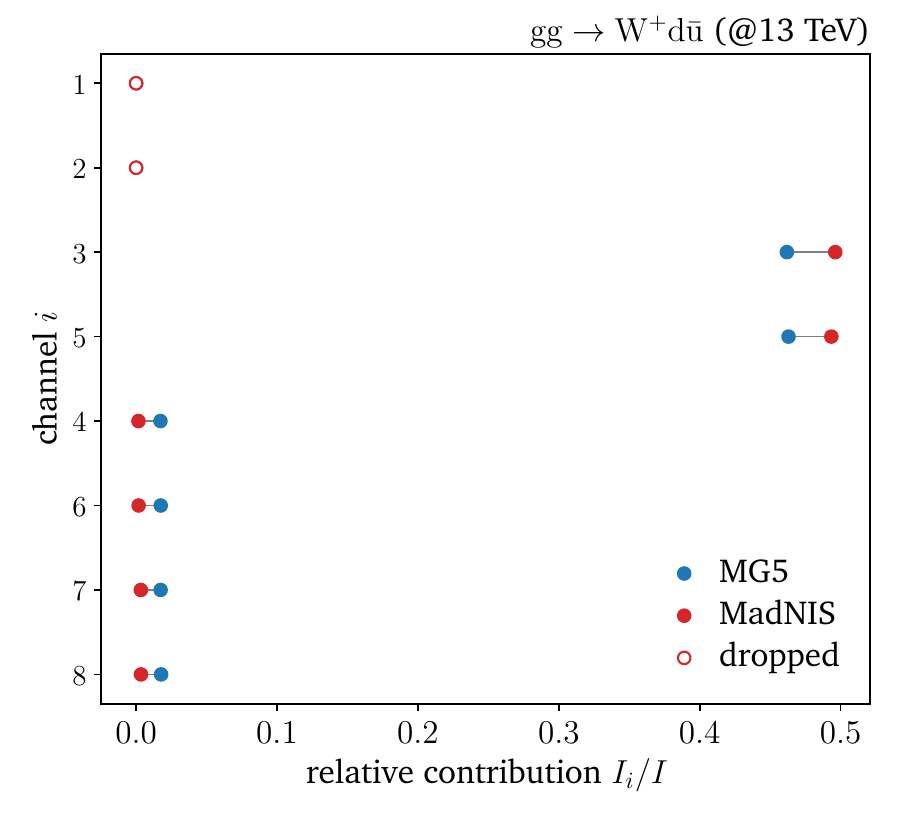}
    \includegraphics[width=0.495\textwidth]{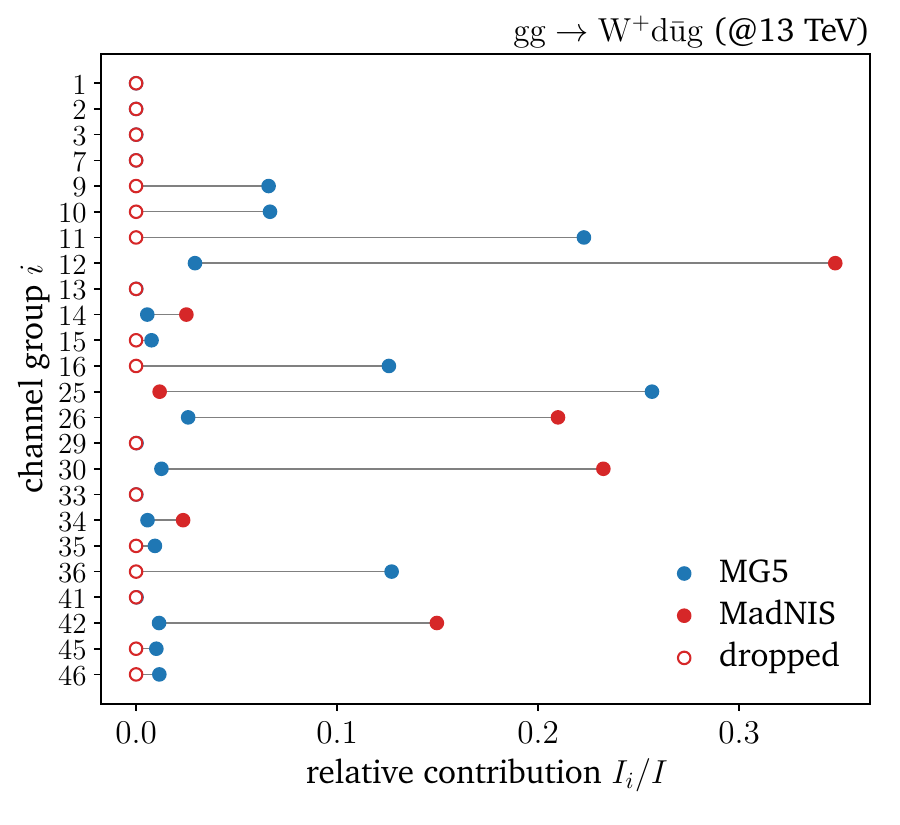}\\
    \includegraphics[width=0.495\textwidth]{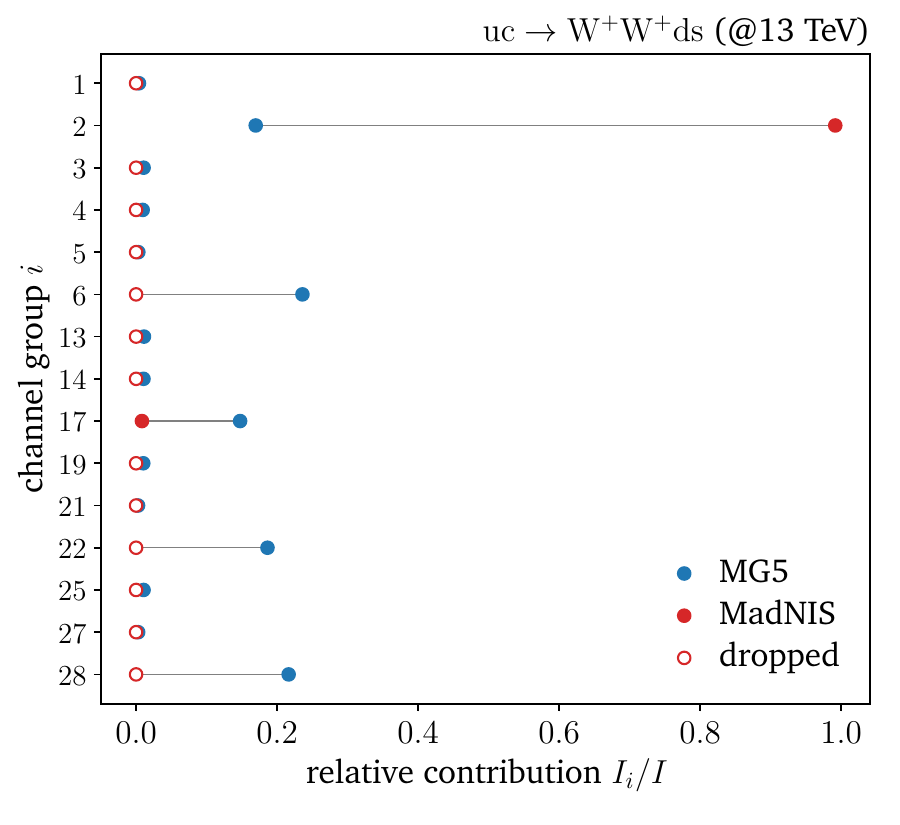}
    \includegraphics[width=0.495\textwidth]{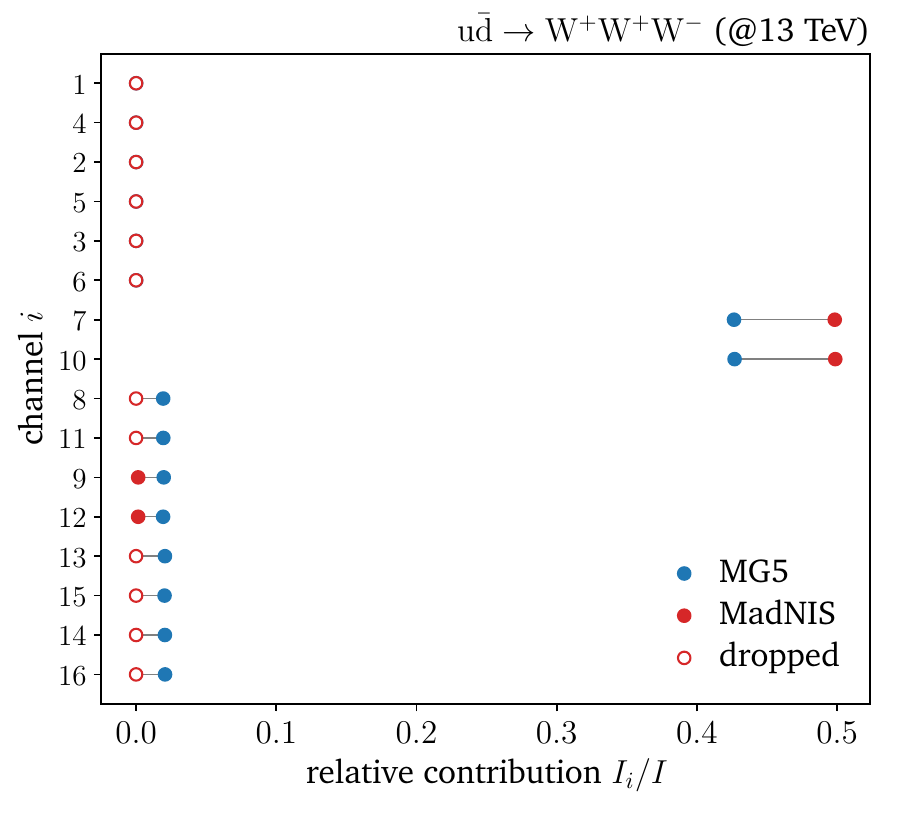}
    \caption{Relative contributions of the channels for W+2jets and
      Triple-W production, and for the channel groups for W+3jets and
      VBS. The channel weights are defined by \mg, their weights are
      learned by \madnis. An empty circle indicates a dropped channel.}
    \label{fig:lollipop}
\end{figure}

It turns our that we can also extract information from the learned
channel weights. We look at the contribution of each channel to the
total cross section,
\begin{align}
  \frac{I_i}{I}=\frac{\int \d^D x\;\ai(x)\,f(x)}{\int \d^D x\;f(x)} \; ,
\end{align}
We find that \madnis changes the contribution of channel groups
coherently, so for W+3jets and Triple-W production we look at the
contributions given by the sum over the channels in the group.

In Fig.~\ref{fig:lollipop}, we show the contribution of \madnis
channels or channel groups, compared to the initial \mg assignments.
We mark dropped channels with empty circles, the number of remaining
active channels corresponds to Tab.~\ref{tab:processes}. We see that
\madnis prefers much fewer channels, illustrating the benefit of our
channel dropping feature.

\begin{figure}
    \centering
    \includegraphics[width=0.30\textwidth]{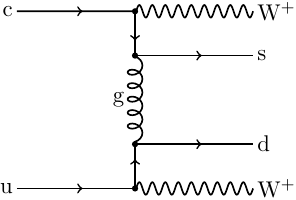}
    \hspace{0.1\textwidth}
    \includegraphics[width=0.30\textwidth]{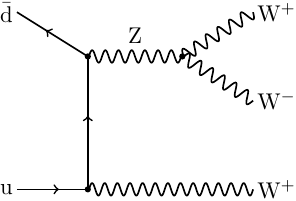}
    \caption{Feynman diagrams corresponding to the dominant channels
      after training \madnis for VBS (left) and Triple-W production
      (right).}
    \label{fig:channel_feynman}
\end{figure}

For VBS and Triple-W production, \madnis adapts the channel weights in
a way that the integrand is almost completely made up from a single
group of symmetry-related channels.  The general behaviour and the
specific choice of channels is consistent between repetitions of the
training. The Feynman diagrams corresponding to these channels are
shown in Fig.~\ref{fig:channel_feynman}.  For VBS five channel groups
significantly contribute to the integral in \mg, all of them with a
$t$-channel gluon or photon.  Of those, \madnis enhances the QCD contribution $\mathcal{O}(\alpha_s^2\alpha^2)$
without an $s$-channel quark propagator. For Triple-W production, one
channel group already dominates the integral in \mg, and it is further
enhanced by \madnis. We give an example for the distribution learned
by the learned channel weights as a function of phase space in
Appendix~\ref{app:cw}.

\subsection{Scaling with jet multiplicity}
\label{sec:implement_scale}

\begin{figure}[b!]
    \includegraphics[width=0.495\textwidth]{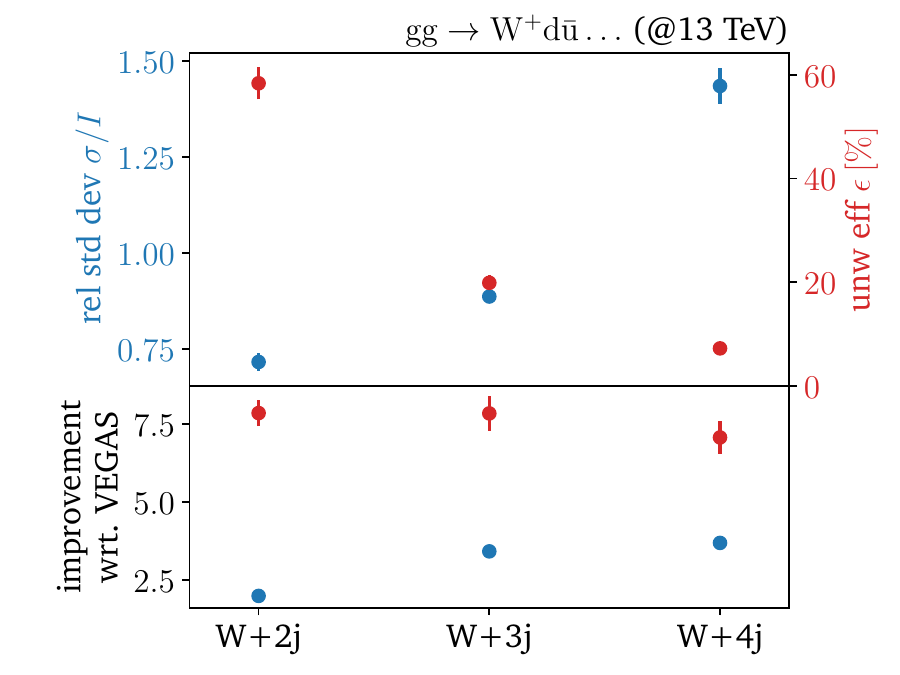}
    \includegraphics[width=0.495\textwidth]{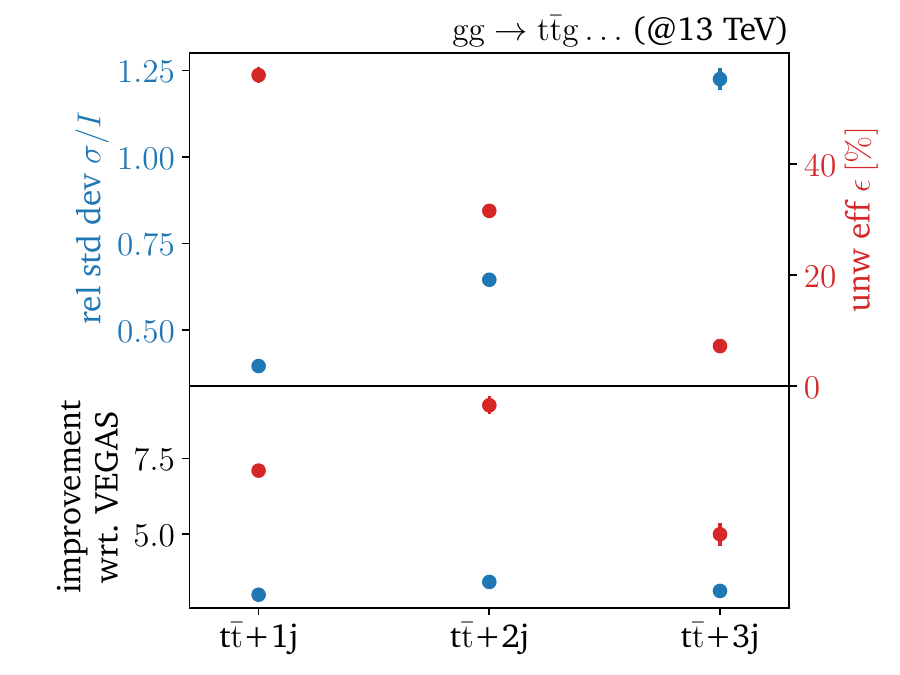}
    \caption{Relative standard deviation and unweighting efficiency
      for W+jets and $\Pt\Ptbar$+jets with different numbers of gluons
      in the final state. The final \madnis performance gain is
      illustrated in the lower panels, just as in Fig.~\ref{fig:process_comparison}.}
    \label{fig:scaling}
\end{figure}

The last challenge of modern event generation \madnis needs to meet is
large number of additional jets.  We study the scaling of the \madnis
performance with the number of gluons in the final state for W+jets
and $\Pt\Ptbar$+jets production. Again, we use the relative standard
deviation $\sigma/I$ and the unweighting efficiency $\epsilon$ as
performance metrics.  As for the final result in
Fig.~\ref{fig:process_comparison}, we train \madnis with all features,
including buffered training with $R_@=5$. The results are shown in
Fig.~\ref{fig:scaling}. While the unweighting efficiency decreases and
standard deviation increases towards higher multiplicities, the gain
over \vegas and \mg remains roughly constant for W+jets
production. For the even more challenging $\Pt\Ptbar$+jets production
the gain decreases for three jets, defining a remaining task for the
final, public release of \madnis.

\section{Outlook}
\label{sec:outlook}

We have, for the first time, shown that modern machine learning leads
to a significant speed gain in \mg. We have
improved the \madnis method~\cite{Heimel:2022wyj} and implemented it
in \mg, to be able to quantify the performance gain from a modern
ML-treatment of phase space sampling. This implementation will allow
us to use the entire \mg functionality while developing an ultra-fast
event generator for the HL-LHC.

Starting from a combined training of a learnable phase space mapping
encoded in an INN and learnable channel weights encoded in a simple
regression network, we have added a series of new features, including
an improved loss function and a fast \vegas initialization. The
combined online and buffered training has been further enhanced by
stratified sampling and channel dropping. The basic structure behind
the phase space mapping is still a state-of-the-art INN with rational
quadratic spline coupling layers, rather than more expressive
diffusion networks~\cite{Butter:2023fov}, because event generation
requires the mapping to be fast in both directions.

The processes we used to benchmark the \madnis performance gain are
W+2,3,4~jets, VBS, Triple-W, and $\Pt\Ptbar$+1,2,3~jets. They combine
large numbers of gauge-related Feynman diagrams with a large number of
particles in the final state. Through \mg, these key processes can
easily be expanded to any other partonic production or decay process.
For the electroweak reference processes we have shown the performance
gain, in terms of the integration error and the unweighting
efficiency, for each of our new features separately. The performance
benchmark was a tuned \mg implementation.  It turned out that learned
phase-space weights, the \vegas initialization, and the stratified
training each contribute to a significant performance gain. Channel dropping and buffered training stabilize the training and limit the
computational cost of the network training. The \madnis gain in the
unweighting efficiency ranges between a factor 5 and a factor 15, the
latter realized for the notorious VBS process.

\section*{Acknowledgements}
OM, FM and RW acknowledge support by FRS-FNRS (Belgian National Scientific Research Fund) IISN projects  4.4503.16. TP would like to thank the Baden-W\"urttemberg-Stiftung for financing through the program \textit{Internationale Spitzenforschung}, project
\textsl{Uncertainties --- Teaching AI its Limits}
(BWST\_IF2020-010). TP is supported by the Deutsche
Forschungsgemeinschaft (DFG, German Research Foundation) under grant
396021762 -- TRR~257 \textsl{Particle Physics Phenomenology after the
Higgs Discovery}. TH is funded by the Carl-Zeiss-Stiftung through the
project \textsl{Model-Based AI: Physical Models and Deep Learning for
Imaging and Cancer Treatment}.
The authors acknowledge support by the state of Baden-Württemberg through bwHPC and the German Research Foundation (DFG) through grant no INST 39/963-1 FUGG (bwForCluster NEMO). Computational resources have been provided by the supercomputing facilities of the Université catholique de Louvain (CISM/UCL) and the Consortium des Équipements de Calcul Intensif en Fédération Wallonie Bruxelles (CÉCI) funded by the Fond de la Recherche Scientifique de Belgique (F.R.S.-FNRS) under convention 2.5020.11 and by the Walloon Region.

\clearpage
\appendix
\section{Hyperparameters}
\label{app:hyper}

\begin{table}[h!]
    \centering
    \begin{small} \begin{tabular}[t]{l|l}
    \toprule
    Parameter & Value \\
    \midrule
    Optimizer & Adam~\cite{Kingma:2014vow} \\
    Learning rate & 0.001 \\
    LR schedule & Inverse decay \\
    Final learning rate & 0.0001 \\
    Batch size & $\min(200 \cdot n_c^{0.8}, 10000)$ \\
    Training length & 88k batches \\
    Permutations & Logarithmic decomposition~\cite{Gao:2020vdv} \\
    Number of coupling blocks & $2 \: \lceil \log_2 D \rceil$ \\
    Coupling transformation & RQ splines~\cite{durkan2019neural} \\
    Subnet hidden nodes & 32 \\
    Subnet depth & 3 \\
    CWnet parametrization & $(\log p_T, \eta,\phi)$ \\
    CWnet hidden nodes & 64 \\
    CWnet depth & 3 \\
    Activation function & leaky ReLU \\
    Max. \# of buffered channel weights & 75 \\
    Buffer size & 1000 batches \\
    Channel dropping cutoff & 0.001 \\
    Uniform training fraction $r$ & 0.1 \\
    \vegas iterations & 7 (7)\\
    \vegas bins & 64 (128) \\
    \vegas samples per iteration & 20k (50k)\\
    \vegas damping $\alpha$ & 0.7 (0.5) \\
    \bottomrule
    \end{tabular} \end{small}
    \caption{\madnis hyperparameters. For the \vegas parameters, the
      first value is used for the pre-training and the value in
      parentheses for the remaining runs.}
    \label{tab:hyperparams}
\end{table}

\section{Channel-weight kinematics}
\label{app:cw}

\begin{figure}[t!]
    \centering
    \includegraphics[width=0.66\textwidth,page=4]{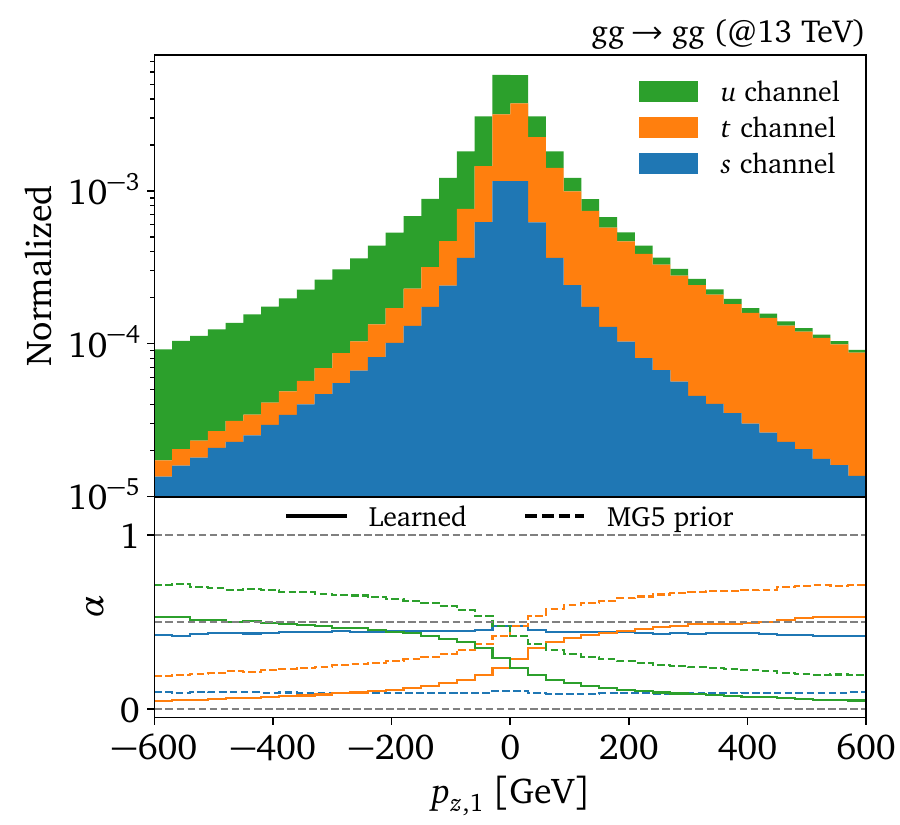}\\
    \caption{Stacked histogram of $z$-component of the first gluon
      momentum for the process $\Pg \Pg \to \Pg \Pg$ generated by
      \madnis. The lower panel shows the bin-wise averages of the
      channels weights as obtained from \mg and the learned channel
      weights.}
    \label{fig:stack}
\end{figure}

To illustrate the learned channel weights we use an example with only
three channels,
\begin{align} 
\Pg \Pg &\to \Pg \Pg\;.
\end{align}
This process has four Feynman diagrams, but \mg only 
constructs 
mappings corresponding to the $s$, $t$ and $u$-channel Feynman
diagrams. The latter two are related by an exchange of the two
final-state gluons and will therefore share the same normalizing flow
in \madnis. We run a \madnis training and use it to generate a set of
weighted events. We keep the channel weights $\alpha^\text{MG}$
provided by \mg and the $\alpha$ obtained from the CWnet. In
Fig.~\ref{fig:stack}, we show a stacked histogram of the $z$ component
of the momentum of the first gluon, because the asymmetric structure
of the $t$ and $u$ channels is best visible for this
observable. Furthermore, we show the average $\alpha^\text{MG}$ and
$\alpha$ in every bin of this histogram. In this example we see that
\madnis does not significantly change the functional form of the
channel weights over phase space, but enhances the contribution of the
$s$ channel, while decreasing the contributions of the $t$ and $u$
channels. This might be because \madnis prefers channels that cover
the entire phase space instead of more specialized channels for simple
processes like this one.

\bibliography{tilman,refs}
\end{document}

%% file: incl_settings.tex
\usepackage[utf8]{inputenc} 
\usepackage[T1]{fontenc} 	
\usepackage[english]{babel} 

\usepackage[bitstream-charter]{mathdesign}
\usepackage{geometry} 		
\usepackage{amsmath} 		
\usepackage{mathtools} 		
\usepackage{float} 			
\usepackage{graphicx} 		
\usepackage{tabularx} 		
\usepackage{booktabs} 		
\usepackage{color, xcolor} 	
\usepackage{pdfpages} 		
\usepackage{extarrows} 		
\usepackage{multirow} 		
\usepackage{multicol} 		
\usepackage{enumitem} 		
\usepackage{xspace} 		
\usepackage{stackrel} 		
\usepackage{tikz} 			
\usepackage{braket} 		
\usepackage{bm} 			
\usepackage{tensor} 		
\usepackage{slashed} 		
\usepackage{siunitx} 		
\usepackage{lastpage} 		
\usepackage{cite} 			
\usepackage[normalem]{ulem} 
\usepackage{fontawesome} 	
\usepackage{tocloft} 		
\usepackage{titlesec} 		
\usepackage{doi} 			
\usepackage{hyperref} 		
\usepackage[most]{tcolorbox} 					
\usepackage[nameinlink, capitalize]{cleveref} 	
\usepackage[nottoc, notlot, notlof]{tocbibind} 	
\usepackage[ruled, vlined]{algorithm2e} 		
\usepackage{makecell}
\usepackage[makeroom]{cancel}
\usepackage{feynmf}

\makeatletter
\def\BState{\State\hskip-\ALG@thistlm}
\makeatother

\makeatletter
\@ifundefined{pdfoutput}{}{\DeclareGraphicsRule{*}{mps}{*}{}}
\makeatother

\makeatletter
\DeclareRobustCommand*{\bfseries}{%
   \not@math@alphabet\bfseries\mathbf
   \fontseries\bfdefault\selectfont
   \boldmath
}
\makeatother

\hypersetup{
	pdftitle={MadNIS},
	pdfauthor={Heimel et al.},
	colorlinks=true, 			
	linkcolor={red!50!black}, 	
	citecolor={blue!50!black}, 	
	urlcolor={blue!80!black} 	
} 

\DeclareSymbolFont{usualmathcal}{OMS}{cmsy}{m}{n}
\DeclareSymbolFontAlphabet{\mathcal}{usualmathcal}

\SetArgSty{textnormal}
\SetKwComment{Comment}{{\small\#}~}{}
\SetCommentSty{mycommfont}

\setitemize{itemsep=0pt, parsep=0pt} 				
\setenumerate{itemsep=0pt, parsep=0pt} 				
\setitemize{itemsep=2pt,topsep=2pt,parsep=0pt,partopsep=0pt,leftmargin=*}
\setenumerate{itemsep=0pt,topsep=2pt,parsep=0pt,partopsep=0pt,labelindent=3pt,leftmargin=*}
\usepackage{amsmath}
 
\DeclareMathOperator\softplus{Softplus}

\usepackage{amsthm} 		
\theoremstyle{definition}

%% file: incl_shortcuts.tex
\definecolor{red_cb}{HTML}{e41a1c}
\definecolor{blue_cb}{HTML}{377eb8}
\definecolor{green_cb}{HTML}{4daf4a}
\definecolor{purple_cb}{HTML}{984ea3}
\definecolor{orange_cb}{HTML}{ff7f00}

\definecolor{EmeraldGreen}{HTML}{1ea78d}
\definecolor{EnglishRed}{HTML}{b02427}
\hypersetup{colorlinks=true,urlcolor=EmeraldGreen,citecolor=EmeraldGreen,linkcolor=EnglishRed}

\newcommand{\ie}{\text{i.e.}\;}

\newcommand{\ai}{\alpha_i}
\newcommand{\aix}{\alpha_{i \xi}}

\newcommand{\Git}{G_{i \theta}}
\newcommand{\git}{g_{i \theta}}
\newcommand{\gbar}{\overline{G}}

\newcommand{\mwith}{\text{with}}
\newcommand{\mand}{\text{and}}
\newcommand{\mfor}{\text{for}}

\newcommand{\qqquad}{\qquad\quad}
\newcommand{\qqqquad}{\qquad\qquad}

\def\d{\mathrm{d}}

\newcommand\one{\leavevmode\hbox{\small1\normalsize\kern-.33em1}}
\newcommand{\imag}{\mathrm{i}} 				

\newcommand{\softmax}{\operatorname{Softmax}}

\newcommand{\loss}{\mathcal{L}} 	

\newcommand{\vegas}{\textsc{Vegas}\xspace}

\newcommand{\mg}{\textsc{MG5aMC}\xspace}
\newcommand{\pythia}{\textsc{Pythia8}\xspace}

\newcommand{\sherpa}{\textsc{Sherpa}\xspace}

\newcommand{\madnis}{\textsc{MadNIS}\xspace}

\newcommand{\whizard}{\textsc{Whizard}\xspace}

\newcommand{\arXiv}[2][]{%
	\ifthenelse{\equal{#1}{}}%
	{\href{http://arxiv.org/abs/#2}{arXiv:#2}}%
	{\href{http://arxiv.org/abs/#2}{arXiv:#2~[#1]}}}

\def\slashchar#1{\setbox0=\hbox{$#1$}           
   \dimen0=\wd0                                 
   \setbox1=\hbox{/} \dimen1=\wd1               
   \ifdim\dimen0>\dimen1                        
      \rlap{\hbox to \dimen0{\hfil/\hfil}}      
      #1                                        
   \else                                        
      \rlap{\hbox to \dimen1{\hfil$#1$\hfil}}   
      /                                         
   \fi}

\newcommand{\tikznode}[2]{%
\ifmmode%
\tikz[remember picture,baseline=(#1.base),inner sep=0pt] \node (#1) {$#2$};%
\else
\tikz[remember picture,baseline=(#1.base),inner sep=0pt] \node (#1) {#2};%
\fi}

\def\mathswitchr#1{\relax\ifmmode{\mathrm{#1}}\else$\mathrm{#1}$\xspace\fi}
\def\mathswitch#1{\relax\ifmmode#1\else$#1$\xspace\fi}

\newcommand{\PWp}{\mathswitchr {W^+}}
\newcommand{\PWm}{\mathswitchr {W^-}}

\newcommand{\Pg}{\mathswitchr g}

\newcommand{\Pd}{\mathswitchr d}
\newcommand{\Pu}{\mathswitchr u}
\newcommand{\Ps}{\mathswitchr s}
\newcommand{\Pc}{\mathswitchr c}

\newcommand{\Pt}{\mathswitchr t}

\newcommand{\Ptbar}{\mathswitchr{\bar t}}

\newcommand{\Pdbar}{\mathswitchr{\bar d}}

\newcommand{\cbox}[1]{\parbox{4em}{\centering {#1}}}